\documentclass[preprint,12pt,authoryear]{elsarticle}

\usepackage{amssymb}
\usepackage{amsmath}
\usepackage{graphicx}% Include figure files
\usepackage{dcolumn}% Align table columns on decimal point
\usepackage{bm}% bold math
\usepackage{hyperref}% add hypertext capabilities
\usepackage{physics}
\usepackage{xr-hyper}
\usepackage{caption}
\usepackage{subcaption}
\usepackage[dvipsnames]{xcolor}

\DeclareMathOperator*{\argmin}{arg\,min}  % for writing argmin

\journal{journal of computational physics}

\begin{document}

\begin{frontmatter}



\title{A fast hybrid classical-quantum algorithm based on block successive over-relaxation for the heat differential equation} %% Article title


\author[azimfarahani]{Azim Farghadan} %% Author name

\author[azimfarahani]{Mohammad Mahdi Masteri Farahani} %% Author name

\affiliation[azimfarahani]{organization={Iranian Quantum Technologies Research Center (IQTEC)},%Department and Organization
            city={Tehran},
            country={Iran}}

\author[akbari]{Mohsen Akbari} %% Author name
%\email[Corresponding Author: ]{mohsen.akbari@khu.ac.ir}
\affiliation[akbari]{organization={Quantum Optics Lab, Department of Physics},%Department and Organization
	addressline={Kharazmi University}, 
	city={Tehran},
	country={Iran}}

%% Abstract
\begin{abstract}
The numerical solution of partial differential equations (PDEs) is essential in computational physics. Over the past few decades, various quantum-based methods have been developed to formulate and solve PDEs. Solving PDEs incur high time complexity for real-world problems with high dimensions, and using traditional methods becomes practically inefficient. This paper presents a fast hybrid classical-quantum paradigm based on successive over-relaxation (SOR) to accelerate solving PDEs. Using the discretization method, this approach reduces the PDE solution to solving a system of linear equations, which is then addressed using the block SOR method. Due to limitations in the number of qubits, the block SOR method is employed, where the entire system of linear equations is decomposed into smaller subsystems. These subsystems are iteratively solved block-wise using Advantage quantum computers developed by D-Wave Systems, and the solutions are subsequently combined to obtain the overall solution. The performance of the proposed method is evaluated by solving the heat equation for a square plate with fixed boundary temperatures and comparing the results with the best existing method. Experimental results show that the proposed method can accelerate the solution of high-dimensional PDEs by using a limited number of qubits up to 2 times the existing method.
\end{abstract}

%% Keywords

\end{frontmatter}



\section{Introduction}
Differential equations are indispensable tools for modeling a wide range of phenomena in physics, engineering, and other applied sciences \cite{article:PDE-application}. Their practical applications extend to numerous industries, where accurate modeling and prediction are crucial for innovation and efficiency \cite{article:PDE-application2}. There are two primary approaches to solving differential equations: analytical methods and numerical methods. However, obtaining an analytical solution for complex systems often presents significant challenges \cite{article:complexity-of-analytical-solution-of-DE}, necessitating the development of numerical methods \cite{article:numerical-sols-DE-nessecary}. One of the most widely recognized numerical methods for solving linear differential equations is the finite difference discretization technique, which involves dividing the domain into cells or volumes \cite{book:discritization1}. This technique transforms linear differential equations into a system of linear equations, \( A \bm{x} = \bm{b} \), where \( A \) is typically a sparse matrix \cite{article:Dwave-linear-system}. As a result of this approach, solutions can be approximated at specific points (center of cells or volumes) within the spatial domain. The most time-consuming part of this method is solving a system of linear equations. The fastest classical algorithm for solving these systems has a time complexity of \( \mathcal{O}(N. \sqrt{\kappa} )\), where \( N \) is the dimension of the matrix \( A \) \cite{book:gussian-elimination} and \(\kappa\)  is its condition number. As the number of cells increases in real-world applications, the size of the linear equation system grows correspondingly, leading to increased computation time.

Quantum algorithms present a promising alternative to classical algorithms, enabling significant reductions in computational complexity \cite{article:HHL, article:shor, article:grover, article:feynman, article:QAOA}. Among these, the Harrow, Hassidim, and Lloyd (HHL) algorithm stands out as a well-known quantum algorithm specifically designed to solve systems of linear equations \cite{article:HHL}. The HHL algorithm achieves a time complexity of \( \mathcal{O}(\text{Poly}(\log(N), \kappa)) \) when the matrix \( A \) is sparse, offering an exponential speedup compared to the fastest classical algorithms. This positions the HHL algorithm as a powerful tool for solving large linear systems. Following its development, researchers have employed the HHL algorithm as a subroutine in solving both ordinary and partial differential equations \cite{article:clader, article:FEM-realize-clader, article:solve-poisson-improve-hhl, article:solve-poisson-improve-hh2, article:bagherimehrab}. However, the output of the HHL-based methods is a quantum state \( \ket{x} \) proportional to the solution, necessitating very resource-intensive quantum tomography to determine the complete solution \( \bm{x} \) \cite{book:nielsen-chuang}. Additionally, the HHL algorithm needs a state preparation step to prepare the quantum state \( \ket{b} \) proportional to \( \bm{b} \) at the input of the algorithm which is also challenging. Moreover, HHL-based methods require extensive quantum resources, such as qubits and gates, making practical application challenging in the era of NISQ (Noisy Intermediate-Scale Quantum) computers \cite{article:HHL-NISQ2}.

An alternative approach for solving a system of linear equations is quantum annealing, where the solution \( \ket{x} \) corresponds to the ground state of the final Hamiltonian \cite{atricle:overview-of-quantum-annealing}. Several studies \cite{article:Ax=b-to-QUBO1, article:Ax=b-to-QUBO4, article:Ax=b-to-QUBO3, article:Ax=b-to-QUBO2} proposed a method to encode \( A\bm{x} = \bm{b} \) to a quadratic unconstrained binary optimization (QUBO) problem through a well-defined protocol and implemented it on a D-Wave quantum annealing system \cite{site:D-Wave-documentation}. Quantum annealing addresses certain challenges associated with HHL-based methods, such as state preparation and tomography. However, the limited number of qubits available on current D-Wave quantum systems (5000+ on the Advantage quantum computer \cite{site:d-wave-advantage-2}) restricts the dimensions of linear systems and, consequently, the practical application of PDEs. In response to this limitation, Pollachini et al. \cite{article:Dwave-linear-system} have adopted a hybrid block Gauss-Seidel method, which decomposes large linear systems into smaller subsystems and solves each subsystem individually, facilitating the solution of large-scale problems with limited quantum resources. Despite its utility, the block Gauss-Seidel method suffers from a suboptimal convergence rate, making it an inefficient iterative approach \cite{book:sor-faster-than-gs3}.

The main purpose of this paper is to solve the steady-state 2D heat equation by a fast hybrid classical-quantum algorithm to decrease computational complexity and increase the convergence rate. Accordingly, this paper proposes a fast paradigm based on the block SOR to accelerate the solution of linear PDEs. This method has superior convergence speed with an optimal choice of over-relaxation parameter \cite{article:block-sor-2}. Finally, the heat equation is implemented in the Advantage system of the D-Wave.

The rest of this paper is organized as follows. Section \ref{sec:propose method} provides a detailed explanation of the proposed approach for solving differential equations, including discussions on finite difference methods, the QUBO formulation of the problem, and the block SOR method. Section \ref{sec:examples} presents an example to illustrate the application of the methodology. Sections \ref{sec: results} and \ref{sec:conclusion} discuss the experimental results and offer concluding remarks, respectively.

\section{Proposed method \label{sec:propose method}}
\begin{figure}[h]
	\centering
	\includegraphics[width=8.6 cm]{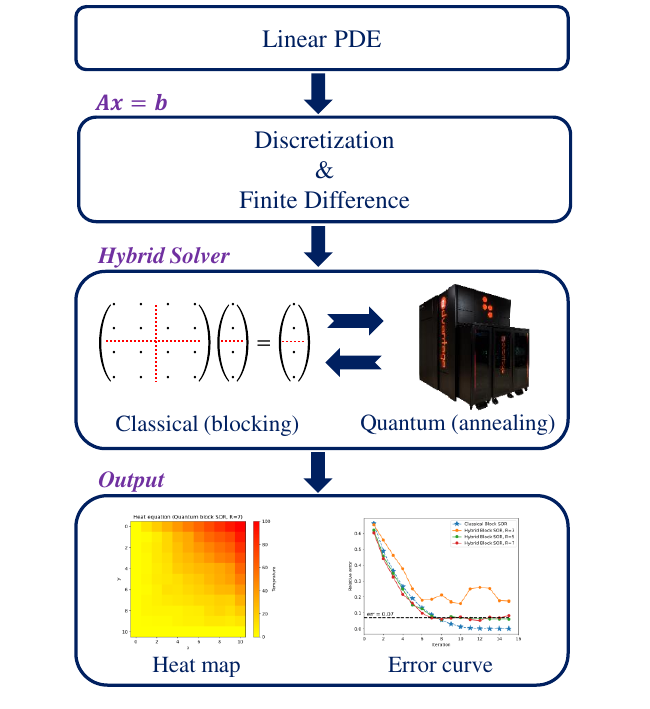}
	\caption{\label{fig3:flowchart} Schematic of the flowchart of proposed method. }
\end{figure}

A description of the proposed algorithm for solving linear PDEs is presented in this section. Linear PDEs are used to model quantum mechanics \cite{article:quantum-mechanics} and various physical processes, such as heat transfer \cite{article:heat-equation}, and wave propagation \cite{article:wave-propagation}. However, solving PDEs for real-world problems with high dimensions remains challenging. Consequently, this paper proposes an efficient hybrid framework based on classical and quantum methods to facilitate this problem. The proposed method follows the sequence shown in FIG. \ref{fig3:flowchart}. A linear PDE is specified as input to the system, and a heat map is generated as output. This algorithm has three main stages. 

First, a system of linear equations (\(A\bm{x} = \bm{b}\)) is generated by discretizing the domain and applying the finite difference method. Secondly, the system of linear equations is fed into the hybrid solver stage. Due to the large size of the system of linear equations in real-world problems, a hybrid solver based on block iterative methods is used. Block iterative methods can decompose large linear equation systems into smaller, more manageable subsystems that can be run on NISQ computers. First, the classical computer decomposes the system of linear equations into smaller subsystems (the blocking stage), and then the quantum system solves these subsystems using the quantum annealing algorithm. When the quantum computer provides solutions, the classical computer consolidates these solutions and calculates the error to determine whether to terminate or continue the iterative process. This process is repeated until convergence of all the subsystems is obtained. As a result of solving these subsystems, we can approximate the desired function at discretized points within the domain.  Finally, a heat map of the solution is generated and an error curve is plotted against the number of iterations to determine the convergence rate.

Each phase of the proposed algorithm, the finite difference discretization technique (Subsection \ref{subsec:Finite Difference}), a fast iterative approach using block SOR (Subsection \ref{subsec:block sor}), and the quantum annealing algorithm for solving a system of linear equations using the QUBO formula (Subsection \ref{subsec: quatnum annealing}) are explained in the following subsections.

\subsection{Finite Difference Method \label{subsec:Finite Difference}}
Solving a differential equation necessitates determining the value of an unknown function at every point within a specified spatial domain. However, in a continuous space, there are an infinite number of such points. To address this challenge, the finite difference discretization technique is employed. This technique discretizes the space by dividing it into a finite number of points and then approximating the derivatives of the unknown function at these discrete points.  The discretization of $\mathbb{R}^2$ can be written as:
\begin{equation}
	x \to x_i, \quad y \to y_j, \ i \in \{1,\cdots, n\}, \ \  j\in \{1,\cdots,m\}.
\end{equation}

The steps in directions $x$ and $y$ can be defined as Eqs. \eqref{eq2:steps in discretization1} and \eqref{eq2:steps in discretization2}.
\begin{eqnarray}
	\label{eq2:steps in discretization1}
	\Delta x &=& \frac{x_n-x_1}{n-1}, \qquad \Delta y = \frac{y_m - y_1}{m-1}\\
	\label{eq2:steps in discretization2}
	x_{i+1} &=& x_i + \Delta x, \qquad y_{j+1} = y_j + \Delta y
\end{eqnarray}

For simplicity, it is assumed that \(\Delta x = \Delta y = h\).  Following the discretization of the spatial domain, we apply the finite difference method, which utilizes the Taylor series expansion to approximate the function values at adjacent points \(x_{i-1} = x_i - h\) and \(x_{i+1} = x_i + h\) as follows: 
\begin{eqnarray}
	\label{eq2:taylor expansion1}
	f(x_i - h) &=& f(x_i) - h\frac{df}{dx}\bigg|_{x_i} + \frac{h^2}{2} \frac{d^2 f}{dx^2}\bigg|_{x_i} - \mathcal{O}(h^3),\ \ \ \ \ \\
	\label{eq2:taylor expansion2}
	f(x_i + h) &=& f(x_i) + h\frac{df}{dx}\bigg|_{x_i} + \frac{h^2}{2} \frac{d^2 f}{dx^2}\bigg|_{x_i} + \mathcal{O}(h^3).
\end{eqnarray}

Accordingly, by subtracting and adding Eqs. \eqref{eq2:taylor expansion1} and \eqref{eq2:taylor expansion2}, the first and second-order derivatives are obtained with an accuracy of \(h^2\) by Eqs. \eqref{eq2:finite difference of first order} and \eqref{eq2:finite difference of second order}, respectively.
\begin{equation}
	\label{eq2:finite difference of first order}
	\begin{split}
		f(x_i + h) - f(x_i - h) = 2h \frac{df}{dx}\bigg|_{x_i} + \mathcal{O}(h^3),\\
		\frac{df}{dx}\bigg|_{x_i} = \frac{f(x_i + h) - f(x_i - h)}{2h} + \mathcal{O}(h^2)
	\end{split}
\end{equation}

\begin{equation}
	\label{eq2:finite difference of second order}
	\begin{split}
		f(x_i + h) + f(x_i - h) = 2f(x_i) + h^2 \frac{d^2 f}{dx^2}\bigg|_{x_i} + \mathcal{O}(h^4),\\
		\frac{d^2 f}{dx^2}\bigg|_{x_i} = \frac{f(x_i + h) - 2f(x_i) + f(x_i - h)}{h^2} + \mathcal{O}(h^2)
	\end{split}
\end{equation}

Given the small value of \(h\),  the terms involving \(h^2\) can be disregarded. As a result, to approximate the partial derivatives of \(f(x,y)\), Eqs. \eqref{eq2:finite difference of partial, first order} and \eqref{eq2:finite difference of partial, second order} must be applied to all discretized points within the domain. 
\begin{equation}
	\label{eq2:finite difference of partial, first order}
	\begin{split}
		\frac{\partial f}{\partial x}(x_i,y_j)\approx \frac{f(x_i + h,y_j) - f(x_i - h,y_j)}{2h},\\
		\frac{\partial f}{\partial y}(x_i,y_j)\approx \frac{f(x_i,y_j + h) - f(x_i,y_j - h)}{2h}
	\end{split}
\end{equation}

\begin{equation}
	\label{eq2:finite difference of partial, second order}
	\begin{split}
		\frac{\partial^2 f}{\partial x^2}(x_i,y_j)\approx \frac{f(x_i + h,y_j) - 2f(x_i,y_j) + f(x_i - h,y_j)}{h^2},\\
		\frac{\partial^2 f}{\partial y^2}(x_i,y_j)\approx \frac{f(x_i,y_j+ h) - 2f(x_i,y_j) + f(x_i,y_j - h)}{h^2}
	\end{split}
\end{equation}

This approximation yields a system of linear equations that, when solved, provide approximate values of the desired function at discretized spatial locations within the domain.

\subsection{Hybrid Solver (Block SOR) \label{subsec:block sor}}
A well-established approach to solving a system of linear equations is the use of block iterative methods \cite{book:iterative-intro}, which improve the solution through a series of iterative approximations. Block iterative methods are useful in numerical linear algebra and computational science when working with partitioned or structured matrices. A block iterative method divides a system of linear equations into smaller blocks and solves each of them separately. The results of all blocks are then combined to produce the overall solution. If the matrix \(A\)=[\(A_{i,j}\)] in the linear system \(A\bm{x} = \bm{b}\) has dimensions \((m \cdot N_b) \times (m \cdot N_b)\), it can be decomposed into blocks as Eq. \eqref{eq1:block form of equations}. Where the blocks \(A_{i,j}\) have size \(m \times m\). 

\begin{equation}
	\label{eq1:block form of equations}
	\left( \begin{matrix}
		\boxed{A_{1,1}} & \boxed{A_{1,2}} & \cdots & \boxed{A_{1,N_b}} \\[10pt]
		\boxed{A_{2,1}} & \boxed{A_{2,2}} & \cdots & \boxed{A_{2,N_b}} \\[10pt]
		\vdots & \vdots & \ddots & \vdots \\[10pt]
		\boxed{A_{N_b,1}} & \boxed{A_{N_b,2}} & \cdots & \boxed{A_{N_b,N_b}}
	\end{matrix} \right)
	\left( \begin{matrix}
		\boxed{\bm{x}_1} \\[10pt]
		\boxed{\bm{x}_2} \\[10pt]
		\vdots \\[10pt]
		\boxed{\bm{x}_{N_b}}
	\end{matrix} \right)
	=
	\left( \begin{matrix}
		\boxed{\bm{b}_1} \\[10pt]
		\boxed{\bm{b}_2} \\[10pt]
		\vdots \\[10pt]
		\boxed{\bm{b}_{N_b}}
	\end{matrix} \right)
\end{equation}

Here, each vector \(\bm{x}_i\) and \(\bm{b}_i\) has dimensions \(m \times 1\). An important aspect of block iterative methods is to convert Eq. \eqref{eq1:block form of equations} into a framework that can be implemented iteratively. As a result, the matrix \(A\) is splitted as $A = E - F$ \cite{article:block-sor-2}. In this equation, \(E\) and \(F\) have block form similar to the matrix \(A\) (i.e., matrices \(E\)=[\(E_{i,j}\)] and \(F\)=[\(F_{i,j}\)] have dimension \((m \cdot N_b) \times (m \cdot N_b)\)) and also \(E\) must be a non-singular matrix. When the system is divided into \( N_b \) blocks, each smaller system has dimensions \(m \times m\) which are compatible with NISQ computers. Consequently, the linear system is transformed into the block iterative form as Eq. \eqref{eq1:iterative the main equation1}:

\begin{equation}
	\label{eq1:iterative the main equation1}
	\sum_{j} E_{i,j} \bm{x}_j^{(k+1)} = \sum_{j} F_{i,j} \bm{x}_j^{(k)} + \bm{b}_i.
\end{equation}
where, \(\bm{x_j}^{(k+1)}\) represents the approximate solution of block \(j\) after \(k+1\) iterations which was computed using the solution of the previous iteration (\(\bm{x_j}^{(k)}\)). The iterative procedure initiates with an initial approximation \(\bm{x_j}^{(0)}\), which may be set to zero, i.e., \(\bm{x_j}^{(0)} = 0\). The critical question then arises: what should the matrices \(E\) and \(F\) in Eq. \eqref{eq1:iterative the main equation1} be? 

Various selections of the matrices \(E\) and \(F\) lead to different block iterative methods. There are advantages and disadvantages to each of these block iterative methods. Since the block SOR iterative method has a high convergence rate, this paper employs it to solve each subsystem. In this way, the convergence rate of the hybrid algorithm for solving linear PDEs is improved. In the block SOR method, the matrices \(E_{i,j}\) and \(F_{i,j}\) are defined as  Eqs. \eqref {eq1:M} and \eqref {eq1:N} respectively.

\begin{eqnarray}
	\label{eq1:M}
	E_{i,j} &=&  \frac{1}{\omega} D_{i,j} + L_{i,j} \ \ \ \ \ \\
	\label{eq1:N}
	F_{i,j} &=&  \left(\frac{1}{\omega} - 1 \right) D_{i,j} - U_{i,j}
\end{eqnarray}
where,
\begin{eqnarray}
	D_{i,j}=\begin{cases}
		A_{i,j}, & \text{if $i=j$}.\\
		0, & \text{otherwise}.
	\end{cases}\\
	U_{i,j} = \begin{cases}
		A_{i,j}, & \text{if $i<j$}.\\
		0, & \text{otherwise}.
	\end{cases}\\
	L_{i,j} = \begin{cases}
		A_{i,j}, & \text{if $i>j$}.\\
		0, & \text{otherwise}.
	\end{cases}
\end{eqnarray}

The parameter \(\omega\) is the over-relaxation parameter, which must be selected optimally to achieve a faster convergence rate. The term "optimal" signifies that the block SOR method achieves its most rapid convergence rate when \(\omega\) is set to \(\omega_{SOR}^{opt}\), is given by \cite{book:iterative-intro}:

\begin{equation}
	\omega_{SOR}^{opt} = \frac{2}{1+\sqrt{1-\rho(H_J)^2}}.
\end{equation}
where \(\rho(H_J)\) denotes the spectral radius (i.e., the maximum absolute value of the eigenvalues) of the Jacobi iteration matrix, which is defined as Eq. \eqref{eq1:spectral radius}. 

\begin{equation}
	\label{eq1:spectral radius}
	H_J = - D^{-1}(L + U)
\end{equation}

The number of iterations required for convergence is not predetermined but is determined by a convergence criterion. Here, we monitor the relative error of the solution vector at the k-th iteration \(\bm{x}^{(k)}\), as a convergence criterion. The relative error is defined as Eq.\eqref {eq1:norm1}. In this equation, \(\bm{x}^{(k)}\) is the approximate solution after \(k\) iterations and \(\bm{x}\) is the exact solution.

\begin{equation}
	\label{eq1:norm1}
	e^{(k)} = \frac{\norm{\bm{x}^{(k)}-\bm{x}}}{\norm{\bm{x}}}
\end{equation}

It is essential to recognize that iterative methods do not universally guarantee convergence for every matrix \( A \). The conditions under which convergence occurs are fundamental to the theoretical framework of iterative methods. Specifically, an iterative method for solving a system of linear equations \( A \bm{x} = \bm{b} \) will converge from any initial guess \( \bm{x}^{(0)} \) if and only if the \textit{spectral radius} of the iteration matrix is less than one \cite{article:convergence-condition-of-iterative}. Consequently, for the block SOR method, we have:

\begin{eqnarray}
	\label{eq1:convergence condition of every iterative method}
	\rho(H_{SOR}) &<& 1, \\
	\label{eq1:H_SOR}
	H_{SOR}(\omega) &=& (D-\omega L)^{-1} \left[ (1-\omega) D + \omega U \right].
\end{eqnarray}

\subsection{Hybrid Solver (Quantum Annealing) \label{subsec: quatnum annealing}}
Obtaining the ground state of a Hamiltonian, particularly in quantum chemistry \cite{ article:ground-state-quantum-chemistry2} and quantum biology \cite{article:ground-state-quantum-biology1}, is an important aspect of quantum mechanics. It is generally not feasible to calculate the ground state of an arbitrary Hamiltonian analytically due to its algebraic complexity \cite{ article:gruond-state-comlpex2}. A popular method for obtaining the ground state is quantum annealing \cite{article:annealing-farhi}. The concept of quantum annealing is derived from the Hamiltonian in Eq. \eqref{eq2:hamiltonian of annealing}. The total evolution time is represented by \(T \). The Hamiltonian is \( H_i \) at \( t=0 \), and at \( t=T \), it transitions to \( H_f \).
\begin{equation}
	\label{eq2:hamiltonian of annealing}
	H(t) = \left(1-\frac{t}{T}\right) H_i + \frac{t}{T} H_f
\end{equation}

The adiabatic theorem \cite{article:adiabatic-theorem-ruski} states that if the evolution time \( T \) is sufficient and the initial state of the system is the ground state of \( H_i \), then the state of the system at time \( T \) will, with a high probability, be \( H_f \). In general, an evolution time \( T \)  is considered sufficient when it is on the order of \( g_{min}^{-2} \), where \( g_{min} \) represents the minimum energy gap between the ground state and the first excited state \cite{article:adiabatic-theorem-ruski, article:adiabatic-theorem}.

D-Wave has successfully implemented quantum annealing based on the Ising Hamiltonian model \cite{article:d-wave-original-idea}. In this model, the initial and final Hamiltonians are expressed as Eqs. \eqref{eq3:initial hamiltonian of d-wave} and \eqref{eq3:final hamiltonian of d-wave}, respectively. In these equations, \( \sigma_x^{(i)} \) and \( \sigma_z^{(i)} \) are spin operators in the \( x \) and \( z \) directions on the \( i \)-th particle, respectively. The coefficients \( \alpha_i \) and \( \beta_{ij} \) are arbitrary real numbers. Since the objective is to find the ground state of the final Hamiltonian, the initial state must be the ground state of the initial Hamiltonian, which is given by $\ket{g} = \ket{+}^{\otimes n}$. The final Hamiltonian represents the Ising model, whose eigenstates can be easily calculated, but the determination of its ground state remains challenging. The primary objective of the D-Wave quantum system is to find the ground state of the Ising model using quantum annealing \cite{article:d-wave-original-idea}.
\begin{eqnarray}
	\label{eq3:initial hamiltonian of d-wave}
	H_i &=& -\sum_{i=1}^n \sigma_x^{(i)}\\
	\label{eq3:final hamiltonian of d-wave}
	H_f &=& \sum_{i=1}^n \alpha_i \sigma_z^{(i)} + \sum_{i,j=1}^n \beta_{ij} \sigma_z^{(i)} \sigma_z^{(j)}
\end{eqnarray}

Moreover, the algebraic structure of the Ising model corresponds to the objective function of QUBO problems. The QUBO problems constitute a significant class of optimization problems, characterized by binary decision variables, a quadratic objective function, and the absence of constraints on the variables \cite{article:QUBO-review}. Consequently, a QUBO formulation can be employed to solve a system of linear equations on a D-Wave quantum system. A linear system \(A\bm{x} = \bm{b}\) can be formulated as an optimization problem by Eq. \eqref{eq3:optimization}.

\begin{equation}
	\label{eq3:optimization}
	\bm{x}_{sol} = \argmin_{\bm{x}} \ (A\bm{x} - \bm{b})^\dagger (A\bm{x} - \bm{b})
\end{equation}

It is clear that the objective function of Eq. \eqref{eq3:optimization} is quadratic, and by expressing the elements of \( \bm{x} \) in binary form, it conforms to the QUBO framework, making it solvable using a D-Wave quantum system. However, the elements $x_i$ of $\bm{x}$ are arbitrary real numbers since it is not possible to exactly represent them in binary form. So, the algorithm starts by converting the real-valued number \( x_i \) into an  \( R \) bits binary format as Eq. \eqref{eq3:R-bit approximation of x_i}. Where, \( x_i \)  is scaled and shifted such that $x_i \in [-d_i,2c_i-d_i)$, where $c_i$ and $d_i$ can be selected depending on the specific of the problem. In this equation \( q^i_r \) is the \( r \)-th bit of \( x_i \) with an \( R \)-bit approximation. When \( d_i > 0\)  and \( c_i > d_i/2 \), the value of \( x_i \) may be positive or negative. This reduces the number of qubits required for a given floating-point accuracy of \( x_i \). There is great benefit in reducing the number of required qubits in the NISQ era systems, where only a few thousand qubits are available.
\begin{equation}
	\label{eq3:R-bit approximation of x_i}
	x_i = c_i \sum_{r=1}^{R} q^i_r  2^{-r}  -d_i
\end{equation}

After substituting \( x_i \) from Eq. \eqref{eq3:R-bit approximation of x_i} into the objective function in Eq. \eqref{eq3:optimization}, the objective function can be rewritten approximately as Eq. \eqref {eq3:QUBO of ax=b in details}.

\begin{equation}
	\label{eq3:QUBO of ax=b in details}
	H = \sum_{r=1}^{R} \sum_{i=1}^{N} \alpha_r^i q^i_r + \sum_{r,s=1}^{R} \sum_{i,j=1}^{N} \beta_{rs}^{ij} q^i_r q^j_s
\end{equation}

In this equation, the coefficients are calculated as Eqs. \eqref {eq3:alpha} and \eqref {eq3:beta}. Here, \( A_{ij} \) denotes the \( ij \)-th element of matrix \( A \) and \( b_{i} \)  are components of the vector \( b \).
\begin{eqnarray}
	\label{eq3:alpha}
	\alpha_r^i &=& -2^{-r+1} \left( \sum_{j,k=1}^{N}  A_{ki} A_{kj} c_i d_j + \sum_{j=1}^{N} A_{ji} c_i b_j \right)\\
	\label{eq3:beta}
	\beta_{rs}^{ij} &=& 2^{-(r+s)} \left( \sum_{k=1}^{N} A_{ki} A_{kj} c_i c_j \right)
\end{eqnarray}

D-Wave has developed a software development interface known as Ocean \cite{site:D-Wave-documentation} to facilitate communication with quantum hardware. Accordingly, the coefficients \(\alpha_i\) and \(\beta_{ij}\) must be provided to the D-Wave Ocean program to solve a system of linear equations. The solution of a system of linear equations on a D-Wave quantum system requires a bit string of \( N \cdot R \) logical qubits. As a result, a substantial number of qubits is required when dealing with large matrices (large \( N \)) or when high accuracy is required (large \( R \)). This emphasizes the importance of hybrid block iterative methods.

Table I shows the comparative resources achieved by the iterative and non-iterative QUBO. The columns of Table I contain the names of the iterative methods, the number of qubits for representing the solution, and run-time respectively. The variable \( T_{N,R} \) in Table 1 represents the time required to run a single QUBO algorithm for solving a linear system of size \( N \), with \( R \) qubits for representing the solution. In contrast, \( T_{N/N_b,R} \) denotes the time needed to solve a system of size \( N/N_b \). The variable \( n_{\text{iter}} \) in the table specifies the number of iterations in the iterative QUBO method which is decreased in our proposed method. Both \( T_{N,R} \) and \( T_{N/N_b,R} \) are dependent on the annealing time and the number of runs necessary to achieve the solutions.

\begin{table}[h]
	\centering
	\caption{Comparison between original QUBO and iterative QUBO methods.}
	\renewcommand{\arraystretch}{1.5} % Adjust the factor to increase row height
	\begin{tabular}{|c|c|c|}
		\hline
		\textbf{Method} & \textbf{Number of Qubits} & \textbf{Time} \\ \hline
		Original QUBO & \( N \cdot R \) & \( T_{N,R} \) \\ \hline
		Iterative QUBO & \( \frac{N}{N_b} \cdot R \) & \( T_{N/N_b,R} \cdot N_b \cdot n_{\text{iter}} \) \\ \hline
	\end{tabular}
\end{table}

\section{ example \label{sec:examples}}
The performance of the proposed method is evaluated using one of the most important PDEs in physics, namely the heat equation. As an example, the heat transfer problem within a given region requires the solution of the heat equation. The heat equation is expressed as Eq. \eqref{eq1:heat equation}.
\begin{equation}
	\label{eq1:heat equation}
	\frac{\partial u}{\partial t} = \alpha^2 \laplacian{u}
\end{equation}

The proposed method is described in detail for a square region of length $L$, with fixed temperatures at the edges. For the stationary case, where the temperature $u$ is independent of time, the heat equation simplifies to Eq. \eqref{eq1:laplace equation}. Where $u = u(x,y,t)$ denotes the temperature at the point $(x,y)$ at time $t$.
\begin{equation}
	\label{eq1:laplace equation}
	\laplacian{u} = 0
\end{equation}

\begin{figure}[h]
	\centering
	\includegraphics[width=8.6 cm]{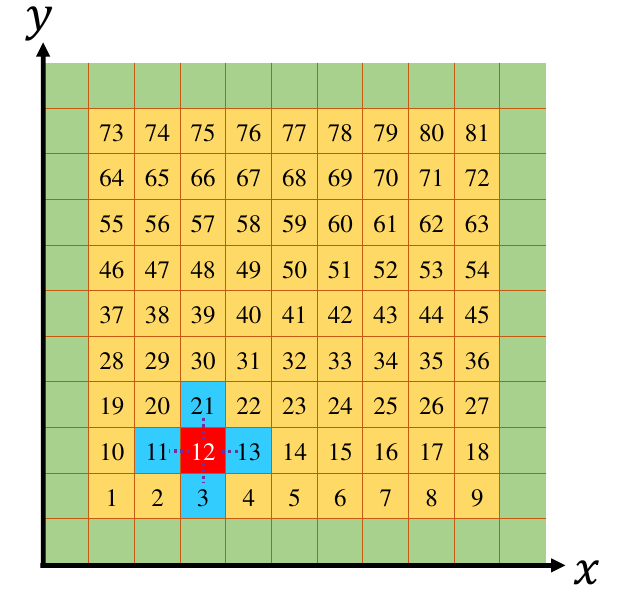}
	\caption{\label{fig2:finite difference of laplacian} The discretization of the spatial domain ($\mathbb{R}^2$) to a grid of cells is depicted. It shows that the approximation of the Laplacian at point 12 requires the function values at points 3, 11, 13, and 21.}
	
\end{figure}

The initial phase of the proposed method involves spatial domain discretization. The region is discretized into a grid consisting of \(11 \times 11\) cells. The discretized spatial points are depicted in FIG. \ref{fig2:finite difference of laplacian}. In this illustration, green points represent the boundaries of the domain, while yellow points represent its interior. Discrete points are considered to be the centers of each cell, where each point corresponds to a number. It is possible to use different numbering schemes for discrete points. As illustrated, we offer snake numbering \cite{book:snake-numbering}. This special numbering yields advantages, such as facilitating access to points during programming. For example, each pair of vertically aligned points has an index difference of 9. Furthermore, this numbering results in a sparse tridiagonal matrix when applied to the finite difference method \cite{book:sor-faster-than-gs3}.

Dirichlet boundary conditions are used to determine the values of a desired function in the boundary. These boundary conditions are defined in Eq. \eqref{eq3:boundary condition}. Based on the known values of temperature at the boundaries, excluding these cells yields a grid of $9 \times 9$ internal cells. Due to this, 81 points of the grid need to be solved for the heat equation.
\begin{equation}
	\label{eq3:boundary condition}
	\begin{split}
		&u(0,y) = u(x,0) = 0\ ^\circ C,\\
		&u(L,y) = \frac{y}{L} \times 100\ ^\circ C,\\
		&u(x,L) = \frac{x}{L} \times 100\ ^\circ C
	\end{split}
\end{equation}

To approximate the heat equation, it is necessary to approximate the Laplacian operator using the finite difference method, as shown in Eq. \eqref{eq2:finite difference of partial, second order}, leading to the approximation given in Eq. \eqref{eq3:laplacian approximation}.
\begin{equation}
	\label{eq3:laplacian approximation}
	\begin{split}
		\laplacian{u} (x_i, y_j) &= \frac{\partial^2 u}{\partial x^2} (x_i, y_j) + \frac{\partial^2 u}{\partial y^2} (x_i, y_j) \\
		\approx \bigg( u(x_i +& h,y_j) + u(x_i - h,y_j)+ u(x_i,y_j + h)  \\
		&\ \ \ \ \ \ + u(x_i,y_j- h) - 4 u(x_i,y_j)\bigg) /h^2
	\end{split}
\end{equation}

It is evident that to approximate the Laplacian operator at a specific point \((x_i, y_j)\), the values of the function at four neighboring points and the point itself are required. For example in FIG. \ref {fig2:finite difference of laplacian}, the Laplacian operator approximation at point 12 is obtained from Eq. \eqref{eq3:example of laplacian approximation}.
\begin{equation}
	\label{eq3:example of laplacian approximation}
	\laplacian{u}\bigg|_{12} \approx \frac{u_3 + u_{11} + u_{13} + u_{21} - 4 u_{12}}{h^2}
\end{equation}

Applying the approximation of the Laplacian operator (Eq. \eqref{eq3:laplacian approximation}) to all interior points within the domain results in a sparse system of linear equations.

\begin{equation}
	A \bm{x} + \bm{b_c} = 0 \quad \implies \quad A \bm{x} = - \bm{b_c}
\end{equation}

In this system, the vector \(\bm{x}\) represents the solution vector \(\bm{x} = \left(u_1, u_2, \cdots, u_{81}\right)^T\), and the vector \(\bm{b_c}\) contains information regarding the boundary function values. The matrix \(A\) has structure as shown in Eq. \eqref{eq3:A of example}.
\begin{equation}
	\label{eq3:A of example}
	A = \begin{pmatrix}
		C & I & 0 & \cdots & 0\\
		I & C & I & \cdots & 0 \\
		\vdots & \ddots & \ddots & \ddots & \vdots \\
		0 & \cdots & I & C & I\\
		0 & \cdots & 0 & I & C
	\end{pmatrix}
\end{equation}

This matrix has a sparse tridiagonal block form, where \(I\) is the identity matrix, and the matrix \(C\) is obtained as in Eq. \eqref{eq3:C and I of A of example}.
\begin{equation}
	\label{eq3:C and I of A of example}
	C =\begin{pmatrix}
		-4 & 1 & 0 & \cdots & 0\\
		1 & -4 & 1 & \cdots & 0 \\
		\vdots & \ddots & \ddots & \ddots & \vdots \\
		0 & \cdots & 1 & -4 & 1\\
		0 & \cdots & 0 & 1 & -4
	\end{pmatrix}
\end{equation}

Consequently, solving the heat differential equation is reduced to solving a system of linear equations, which is then addressed using the block SOR method.

\section{Discussion and Experimental results \label{sec: results}}

The performance of the proposed method has been evaluated by solving the heat equation for a square plate with fixed temperatures, using the Advantage D-Wave system. To assess efficiency and accuracy, the problem is numerically addressed through the finite difference method, partitioning the domain into \( N_b \) segments of equal length. This approach allows for solving larger linear systems of equations, thus overcoming the constraints posed by the limited number of qubits in D-Wave quantum processing units.

The validation of the proposed method is integrated with the initial implementation of the block Gauss-Seidel method, as suggested by \cite{article:Dwave-linear-system}, due to its close resemblance to our proposed algorithm and its proven effectiveness in addressing challenges associated with high-dimensional problems. To address the limitations of the block Gauss-Seidel method, particularly its slow convergence, our proposed method was introduced, demonstrating faster convergence with fewer iterations and reducing overall execution time. The iterative process in the block SOR method is conceptually analogous to that of the block Gauss-Seidel method; for \(\omega=1\), the block SOR method transforms into the block Gauss-Seidel method. 

\begin{figure} [h]
	\centering
	\begin{subfigure}[b]{8.6 cm}
		\centering
		\includegraphics[width=\textwidth]{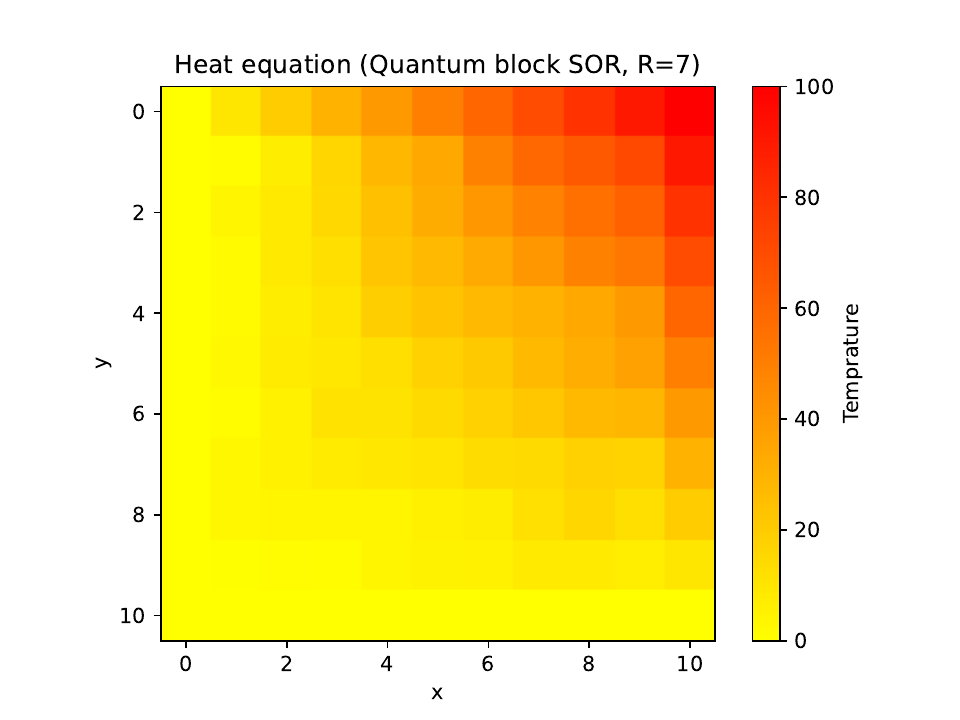}
	\end{subfigure}
	\caption{The solution of heat equation with boundary condition in Eq. \eqref{eq3:boundary condition} with hybrid block SOR algorithm in the square plate.}
	\label{fig:heat-sols}
\end{figure}

The heat equation example demonstrates the effectiveness of the proposed methodology, with three studies. These studies examine the effect of the number of qubits for representing the solution. The convergence criterion was selected in alignment with the approach by \cite{article:Dwave-linear-system} (relative error = 0.08).  FIG. \ref {fig:heat-sols} shows the temperature distribution in the square plate after convergence. 

\begin{figure}[h]
	\centering
	\begin{subfigure}[b]{0.48\textwidth}
		\centering
		\includegraphics[width=\textwidth]{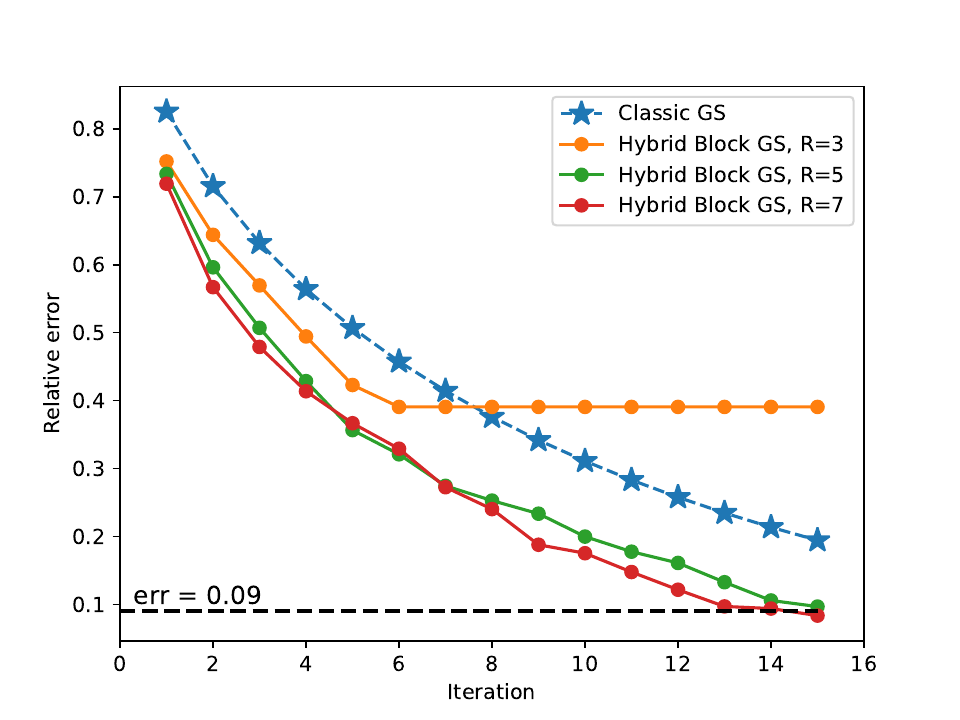}
		\caption{Classic and hybrid block Gauss-Seidel method \cite{article:Dwave-linear-system}.}
		\label{fig:bGS-classical-quatnum}
	\end{subfigure}
	%	\hfill
	\begin{subfigure}[b]{0.48\textwidth}
		\centering
		\includegraphics[width=\textwidth]{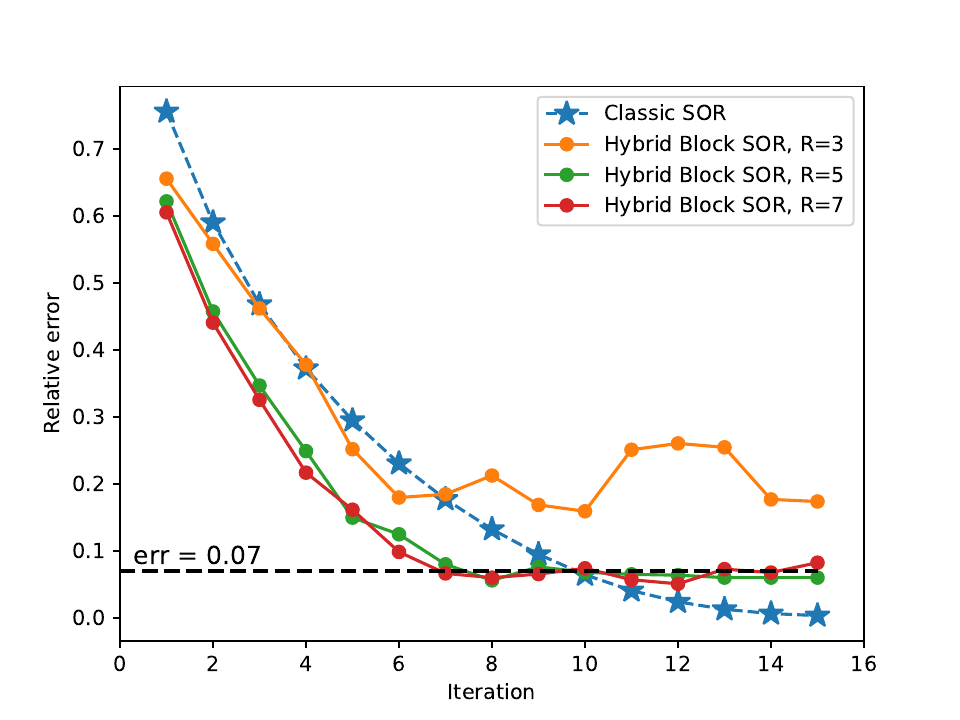}
		\caption{Classic block SOR and hybrid block SOR method.}
		\label{fig:bSOR-classical-quatnum}
	\end{subfigure}
	%	\hfill
	\begin{subfigure}[b]{0.48\textwidth}
		\centering
		\includegraphics[width=\textwidth]{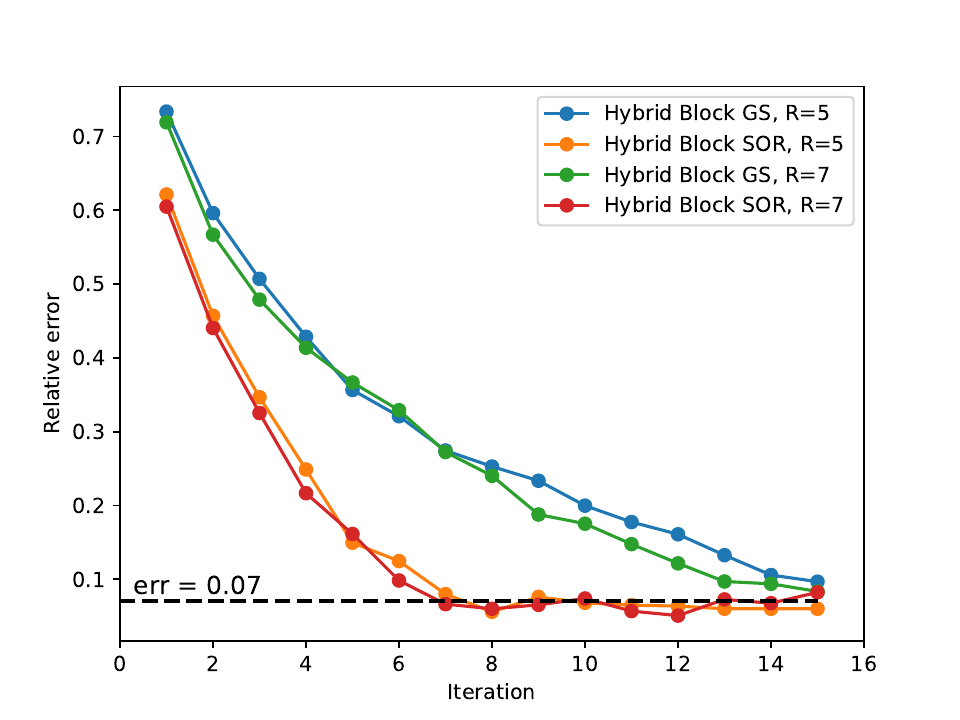}
		\caption{Hybrid block Gauss-Seidel and block SOR method.}
		\label{fig:bGS-classical-SOR-quantum}
	\end{subfigure}
	\caption{The relative error is analyzed as a function of the number of iterations for the Block Gauss-Seidel and Block SOR methods, and a comparison of them, considering varying qubit counts in the numerical representation of the solution vector \( x \).}
	\label{fig:iterative methods of block GS and SOR}
\end{figure}

FIG. \ref {fig:iterative methods of block GS and SOR} illustrates the relationship between the relative error, as defined in Eq. (24), and the number of iterations of the Block Gauss-Seidel and Block SOR methods, comparing their performance across varying quantities of qubits used in the numerical representation of the solution. We discussed the influence of the number of qubits for representing the solution (set to 3, 5, and 7) on the convergence rate.  This example shows the minimum error when seven qubits are utilized within the Advantage system. 

The experimental results indicate that the proposed method outperforms the block Gauss-Seidel method in all studied scenarios. The results show that our hybrid quantum-classical method accelerates the solution of PDEs by up to two times compared to the block Gauss-Seidel method. While the Gauss-Seidel method achieves an accuracy of 0.09 after 15 iterations, the proposed method reaches the desired accuracy by the 7th iteration. Unlike in classical algorithms, we observed that the error does not consistently decrease, partly due to the limited floating-point accuracy of the variables \( x_i \), which prevents further improvement after a certain number of iterations. Moreover, the relative error reduces as the number of qubits for representing the solution increases, since \(R\) determines the significant figures of the elements in \(\bm{x}\).

\section{Conclusion \label{sec:conclusion}}

This paper presents a fast hybrid classical-quantum approach to solving PDEs using a combination of quantum annealing and the block SOR method, which significantly reduces the computational complexity of solving PDEs. The proposed method addresses the challenge of high-dimensional PDEs, which are computationally intensive to solve using traditional methods, by leveraging the capabilities of quantum computing within a hybrid framework. Through finite difference discretization, the solution of PDEs is reduced to the solution of a system of linear equations, which is then solved using the block SOR method. This block-wise decomposition allows for the efficient handling of larger systems by partitioning them into smaller subsystems that can be processed within the constraints of current quantum hardware. The quantum component of the approach, implemented on D-Wave's Advantage quantum computers, solves these systems at a subsystem level, while the classical component integrates the solutions to achieve the overall result.

The performance of the proposed hybrid method was evaluated by solving the heat equation for a square plate with fixed boundary temperatures. The experimental results indicate that the proposed hybrid quantum-classical method is a promising approach to solving PDEs more efficiently, particularly in high-dimensional scenarios. The method not only accelerates the computation but also achieves comparable accuracy to classical methods, highlighting its potential for practical applications in computational physics and beyond.

The proposed method enables a substantial reduction in the number of logical qubits, from \( N \cdot R \)  to \( N \cdot R/ N_b \), though this reduction comes at the cost of an increased number of iterations required to achieve the desired accuracy. Our results demonstrate that the block SOR method significantly outperforms the block Gauss-Seidel method, achieving convergence up to two times faster. It was observed that increasing the number of qubits used in the numerical representation of the unknown variables enhances the accuracy of the solution.


\end{document}